\documentclass[aps,prl,
twocolumn,nofootinbib]{revtex4-1}
\usepackage{graphicx,epsf,amssymb,amsbsy,amsfonts,amssymb,amsmath,mathtools}
\usepackage[hidelinks,breaklinks]{hyperref}
\usepackage[dvipsnames]{xcolor}
\usepackage{tikz-cd}
\usepackage[normalem]{ulem}

\newcommand{\eq}{\begin{equation}}
\newcommand{\eqe}{\end{equation}}

\newcommand{\eqa}{\begin{eqnarray}}
\newcommand{\eqae}{\end{eqnarray}}

\def\tr{\text{tr}} 

\def\R{\mathbb{R}}

\def\<{\langle}
\def\>{\rangle}
\def\+{\dagger}

\newcommand{\SU}[1]{\text{SU}(#1)}
\newcommand{\U}[1]{\text{U}(#1)}
\newcommand{\normord}[1]{:\mathrel{#1}:}
\newcommand{\sixj}[6]{ \begin{Bmatrix}
		#1 & #2 & #3 \\
		#4 & #5 & #6 
\end{Bmatrix}}

\begin{document}

\title{Nonrelativistic Corners of \texorpdfstring{${\cal N} = 4$}{N=4}
	Supersymmetric Yang--Mills Theory}

\author{Troels Harmark}
\email{harmark@nbi.ku.dk}

\author{Nico Wintergerst}
\email{nico.wintergerst@nbi.ku.dk}

\affiliation{Niels Bohr Institute, Copenhagen University\\
Blegdamsvej 17, DK-2100 Copenhagen \O, Denmark}

\begin{abstract}
\noindent We show that ${\cal N} = 4$ supersymmetric--Yang--Mills (SYM) theory on $\R \times S^3$ with gauge group $\SU{N}$ is described in a near-BPS limit by a simple lower-dimensional nonrelativistic field theory with $\SU{1,1} \times \U{1}$ invariant interactions.
In this limit, a single complex adjoint scalar field survives, and part of its interaction is obtained by exactly integrating out the gauge boson of the SYM theory. Taking into account normal ordering, the interactions match the one-loop dilatation operator of the $\SU{1,1}$ sector, establishing the consistency of the limit at the quantum level.
We discover a tantalizing field-theoretic structure, corresponding to a $(1+1)$-dimensional complex chiral boson on a circle coupled to a nondynamical gauge field, both in the adjoint representation of $\SU{N}$. 
The successful construction of a lower-dimensional nonrelativistic field theory in the $\SU{1,1}$ near-BPS limit provides a proof of concept for other BPS bounds. These are expected to lead to richer field theories in nonrelativistic corners of ${\cal N} = 4$ SYM that include fermions, gauge fields and supersymmetry and can provide a novel path towards understanding strongly coupled finite-$N$ dynamics of gauge theories.
\end{abstract}

\maketitle

\section*{Introduction}

Through the AdS/CFT correspondence, type IIB string theory on an AdS$_5 \times S^5$ background is conjectured to possess a dual description in terms of ${\cal N} = 4$ super Yang--Mills (SYM) theory with gauge group $\SU{N}$ and coupling $g$. 
In principle, solving the gauge theory 
would provide the full dynamics of strings and thus reveal the emergence of gravity and black holes from a quantum theory. In practice, this daunting task calls for a more feasible approach.
One possibility is to take the planar limit $N\rightarrow \infty$ while keeping the 't Hooft coupling $\lambda = g^2 N$ fixed. Here, one can find the full spectrum by employing a beautiful integrable structure  \cite{Beisertandothers2012}. 
Another possibility is to explore the theory at weak 't Hooft coupling, while keeping $N$ finite  \cite{Sundborg2000,AharonyMarsanoMinwallaPapadodimasVanRaamsdonk2004}. 
Either approach, however, has important limitations. 
In the planar limit, the geometry is fixed and gravity can at best be taken into account perturbatively through $1/N$ corrections. Regimes of strong gravity, in particular black holes, become inaccessible. At weak coupling, on the other hand, finite $N$ contributions are simpler to compute, but the dual string theory ceases to be geometrical, at least in the semiclassical sense. 

In this letter, we explore an alternative idea. In a nonrelativistic limit of the AdS/CFT correspondence \cite{GomisGomisKamimura2005,BagchiGopakumar2009,HarmarkHartongObers2017,HarmarkHartongMenculiniObersYan2018}, both strong dynamics of gravity and a semiclassical geometry 
can be retained, but the quantum field theory side may still simplify sufficiently to enable a direct quantitative study of its strongly coupled finite-$N$ regime.

We take a major step in this direction by considering near-BPS corners of ${\cal N} = 4$ SYM in which the dynamics becomes explicitly nonrelativistic.
 At the hand of a concrete example, we demonstrate
that ${\cal N} = 4$ SYM on a three-sphere 
close to a particular BPS bound is effectively described by the Hamiltonian of a lower-dimensional nonrelativistic field theory. Only a subset of the degrees of freedom contribute and an emergent $\U{1}$ global symmetry corresponds to the conservation of particle number, in accordance with the nonrelativistic nature of the theory. In addition, the interactions are invariant under an additional global $\SU{1,1}$ symmetry, characterizing the concrete bound we are considering.

The BPS bound considered in this letter is 
\begin{equation}
\label{su11_bound}
E \geq Q_1 + S_1
\end{equation}
where $E$ is the energy, $Q_1$ one of the R-charges and $S_1$ one of the angular momenta of ${\cal N}=4$ SYM on a three-sphere. We explore the
 near-BPS limit
\begin{equation}
\label{eq:limit}
\lambda \to 0,\ \ \text{ with }\ \ \frac{E-Q_1-S_1}{\lambda} \ \ \text{ finite, }\ \ N\text{ fixed.}
\end{equation}
That this type of limit, known as a Spin Matrix theory limit, reveals nontrivial dynamics close to BPS bounds was discovered and examined in \cite{HarmarkOrselli2014}. In this work, we find the first clear evidence of a nonrelativistic field-theoretic structure emerging from the near-BPS limit associated with \eqref{su11_bound}. Importantly, we chose the specific bound \eqref{su11_bound} mainly to provide an accessible representative for a proof of concept. All major subtleties of the constructions are captured, allowing one to
readily generalize our methods to other near-BPS limits.

\begin{figure}[h!]
\includegraphics{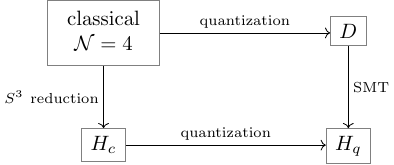}
\caption{Commutative diagram illustrating that quantizing ${\cal N}=4$ SYM at one-loop and subsequently taking a near-BPS limit is equivalent to considering the same limit on an $S^3$ and quantizing the resulting effective theory.}
\label{fig:commd}
\end{figure}

We can illustrate our approach by the commutative diagram displayed in Fig.~\ref{fig:commd}. 

On the one hand, we can take the limit by considering the classical field theory of ${\cal N} = 4$ SYM compactified on a three-sphere. We do this in the first part of this work, which includes a nontrivial contribution to the interactions from integrating out the gauge field degrees of freedom. In this way, we find a classical Hamiltonian $H_{\rm c}$ that describes a lower-dimensional nonrelativistic field theory. Quantizing it and deriving nontrivial normal-ordering terms, leads to the quantum Hamiltonian $H_{\rm q}$.

On the other hand, one can start by quantizing ${\cal N}=4$ SYM.
The spectrum of two-point functions is described by the dilatation operator $D$ \cite{MinahanZarembo2003,BeisertKristjansenPlefkaStaudacher2003,BeisertKristjansenStaudacher2003}.
One can then follow the Spin Matrix theory limit procedure of \cite{HarmarkOrselli2014} with $D$ as starting point, and  take the near-BPS limit in which
only the one-loop contribution to $D$ survives, and the Hilbert space reduces to the $\SU{1,1}$ subsector. We show in this letter that this matches perfectly with $H_{\rm q}$. This implies that the highly nontrivial quantum field theory computation that leads to the relevant part of the one-loop dilatation operator \cite{Beisert2004}, are captured by simple normal-ordering contributions to our classical Hamiltonian $H_{\rm c}$. 

\section*{Classical theories from sphere reduction}

The first step towards closing the diagram shown in Fig.~\ref{fig:commd} is by reducing classical ${\cal N} = 4$ SYM in the near-BPS limit on the three-sphere.
We consider the bound \eqref{su11_bound} in the purely bosonic sector, 
and set all fermion fields to zero. Moreover, we introduce complex combinations of the real scalar fields that transform in the ${\bf 6}$ of $\SU{4}$, $\Phi_a \equiv \phi_{2a - 1} + i\phi_{2a}$, with $a = 1,2,3$. Canonically normalizing the gauge field, the relevant part of the Lagrangian on the three-sphere with unit radius is
\begin{multline}
\label{eq:n4lag}
L = \int_{S^3}\tr\Bigg\{-\frac{1}{4} F_{\mu\nu}^2 - |D_\mu \Phi_a|^2 - |\Phi_a|^2\\
- \frac{g^2}{2} \sum_{a,b}\left(|[\Phi_a,\Phi_b]|^2 + |[\Phi_a,\bar\Phi_b]|^2\right)\Bigg\}\,,
\end{multline}
where bars denote hermitian conjugation. Here, $F_{\mu\nu} = \partial_\mu A_\nu - \partial_\nu A_\mu + ig[A_\mu,A_\nu]$, $D_\mu\Phi_a = \partial_\mu \Phi_a + ig [A_\mu,\Phi_a]$ and both the gauge field $A_\mu$ and the scalars transform in the adjoint representation of $\SU{N}$. 
From Eq.~\eqref{eq:n4lag} we derive the Hamiltonian and determine the relevant propagating degrees of freedom 
 from the quadratic part alone.
  To this end, we adopt Coulomb gauge in order to eliminate unphysical degrees of freedom but
   keep track of the interactions between scalars that are mediated by the longitudinal and temporal gluons. As we will see, taking the limit \eqref{eq:limit} allows us to explicitly integrate out even the transverse gluons, giving rise to an effective theory for a single complex scalar.

Fields are decomposed into spherical harmonics on the $S^3$, as reviewed in detail e.g. in \cite{IshikiTakayamaTsuchiya2006}. 
Scalars are written as $\Phi_a = \sum_{J,M} \Phi_a^{JM} {\cal Y}_{JM}$, while vectors decompose into vector spherical harmonics as $A_i = \sum_{JM} \sum_{\rho=-1}^{1} A^{JM}_{(\rho)} {\cal Y}_{JM\rho,i}$, with $\rho = 0,\pm 1$ labeling the longitudinal and transverse harmonics, respectively. Here, $M \equiv (m,\tilde{m})$, with $m$ and ${\tilde m}$ running from $-J$ to $J$ for scalar spherical harmonics. For vectors, they run from $-Q$ to $Q$ and $-\tilde{Q}$ to $\tilde{Q}$, respectively, where $Q = J + (1+\rho)\rho/2$ and $\tilde{Q} = J - (1-\rho)\rho/2$. Since the harmonics ${\cal Y}_{JM\pm1,i}$ are transverse, the Coulomb gauge condition $\nabla^i A_i$ reduces to $A^{JM}_{(\rho = 0)} = 0$, and both the temporal and longitudinal gluon can be directly integrated out. The resulting Hamiltonian is
\begin{align}
\label{eq:ham}
H &= \tr\sum_{J,M} \Bigg[\frac{1}{2}\left( |\Pi_{(\rho)}^{JM}|^2 +  \omega_{A,J}^2 |A_{(\rho)}^{JM}|^2\right)\nonumber\\
&+ |\Pi_a^{JM}|^2 + \omega_J^2 |\Phi_a^{JM}|^2
+ \frac{1}{8J(J+1)}|{j}_0^{JM}|^2\nonumber\\
&- 4g\sum_{J_i,M_i}\sqrt{J_1(J_1+1)} {\cal D}^{J_2M_2}_{J_1M_1, JM\rho}A^{JM}_{(\rho)}[\Phi_a^{J_1M_2},\bar\Phi_a^{J_2M_2}] \nonumber \\
&+ \frac{g^2}{2}\Bigg|\sum_{J_i,M_i}{\cal C}^{J_2M_2}_{J_1M_1,J M}[\Phi_a^{J_1M_1},\bar\Phi_a^{J_2M_2}]\Bigg|^2\Bigg]\,,
\end{align}
with the scalar charge density
\begin{multline}
j_0^{JM} = i g \sum_{J_i,M_i} {\cal C}^{J_2M_2}_{J_1M_1,JM}\\
\times \left([\bar\Phi_a^{J_2M_2},\bar\Pi_a^{J_1M_1}] + [\Phi_a^{J_1M_1},\Pi_a^{J_2M_2}]\right)\,.
\end{multline}
Here, $\Pi_a$ is the momentum conjugate to $\Phi_a$ and $\Pi_{(\rho)}$ to $A_{(\rho)}$, $\omega_J\equiv 2J+1$, $\omega_{A,J}\equiv 2J+2$ and bars denote hermitian conjugation. 
Doubly occurring indices $a$ and $\rho = \pm 1$ are summed over. The coefficients ${\cal C}$ and ${\cal D}$ are Clebsch--Gordan coefficients that couple three scalar, or one scalar and its derivative with a vector spherical harmonic, respectively,
derived for example in \cite{HamadaHorata2003,IshikiTakayamaTsuchiya2006}. They read
\begin{multline}
{\cal C}^{JM}_{J_1M_1\,J_2M_2} = \sqrt{\frac{(2J_1+1)(2J_2+1)}{2J+1}}\\
\times C^{Jm}_{J_1m_1\,J_2m_2}C^{J\tilde{m}}_{J_1\tilde{m}_1\,J_2\tilde{m}_2}\,,
\label{eq:C_exp}
\end{multline}
\begin{multline}
{\cal D}^{JM}_{J_1M_1\,J_2M_2\rho_2} = (-1)^{-\frac{1}{2}+J+J_1 + J_2}(2J_1+1)\\
\times\sqrt{\frac{(2J_2+1)(2J_2 + 3)}{2J+1}}\sixj{J_1}{J_1}{1}{J_2 - \frac{\rho-1}{2}}{J_2 + \frac{\rho+1}{2}}{J}
\label{eq:D_exp}
\end{multline}
where $\{\}$ is a Wigner $6$-$j$ symbol, $C^{Jm}_{J_1m_1\,J_2m_2}$ are ordinary $\SU{2}$ Clebsch--Gordan coefficients and the latter expression is explicitly valid for $\rho = \pm 1$.
The first interaction in $H$ is the Coulomb term between scalar charges, while the other two are the scalar--gluon, and scalar--scalar interaction present in the ${\cal N}=4$ Hamiltonian. The theory is supplemented with a gauge singlet constraint that arises from integrating out the $(J,M) = (0,0)$-mode of the temporal gauge field. 

The rotation generator $S_1$ reads
\begin{multline}
S_1 = i \sum_{JM} (\tilde{m} - m) \tr\bigg(\Phi_a^{JM}\Pi_a^{JM} - \bar{\Phi}_a^{JM}\bar{\Pi}_a^{JM} \\
+ \frac{1}{2} \left( {A}_{(\rho)}^{JM}\Pi_{(\rho)}^{JM} - \bar{A}_{(\rho)}^{JM}\bar{\Pi}_{(\rho)}^{JM}\right)\bigg)\,,
\end{multline}
while the relevant $\SU{4}$ R-charge is given by
\begin{equation}
Q_1 = i \sum_{JM} \tr(\Phi_1^{JM}\Pi_1^{JM} - \bar{\Phi}_1^{JM}\bar{\Pi}_1^{JM})\,.
\end{equation}
The propagating degrees of freedom can be deduced by demanding $H - S_1 - Q_1 = {\cal O}(g)$ for $g \to 0$, which in particular requires the ${\cal O}(g^0)$-contributions to vanish. Defining $\Delta m \equiv m - \tilde{m}$, we obtain for these 
\begin{multline}
H - S_1 - Q_1|_{g = 0} =  \tr\sum_{JM}\Bigg(
\frac{1}{2}\left(\left|\Pi_{(\rho)}^{JM} - i\Delta m\bar{A}_{(\rho)}^{JM}\right|^2
\right. \\ \left. + (\omega_{A,J}^2 - \Delta m^2) |A_{(\rho)}^{JM}|^2\right)
+ |\Pi_a^{JM} + i(\delta^a_{1}-\Delta m)\bar\Phi_a^{JM}|^2 \\
+ (\omega_J^2 - (\delta^a_{1}-\Delta m)^2) |\Phi_a^{JM}|^2\Bigg)\,.
\end{multline}
Given the form of $\omega_J$ and $\omega_{A,J}$, it is not hard to derive the following set of constraints on $A_{(\rho)}$ and two of the scalar fields:
\begin{equation}
\label{eq:APi_zero}
\begin{array}{c}\displaystyle
\Phi_2 = \Phi_3 = \Pi_2 = \Pi_3 = {\cal O}(g)\,,
\\[3mm] \displaystyle
A_{(\rho)}^{JM} = {\cal O}(g)\,,\quad\Pi_{(\rho)}^{JM}-i\Delta m\bar{A}_{(\rho)}^{JM} = {\cal O}(g)\,.
\end{array}
\end{equation}
For $\Phi_1$, one finds for 
$J = -m = \tilde{m}$
\begin{equation}
\label{eq:Phi_const}
\Pi_1^{J,-J,J} + i\omega_J\bar\Phi_1^{J,-J,J} = {\cal O}(g)\,,
\end{equation}
and for all other $m$, $\tilde{m}$
\begin{equation}
\label{eq:Phizero_const}
\Phi_1^{JM} = \Pi_1^{JM} = {\cal O}(g)\,.
\end{equation}
Each of these constraints eliminates a propagating degree of  freedom; the right hand sides of Eqs.~\eqref{eq:APi_zero}-\eqref{eq:Phizero_const} depend on the field equations and can be deduced by demanding consistency with the full Hamiltonian evolution. All of these vanish, except for $A^{JM}_{(\rho)}$, since it is the only field that appears linearly in Eq.~\eqref{eq:ham}. There, one obtains
\begin{multline}
\label{eq:A_const}
A_{(\rho)}^{JM} = \sum_{J_i,M_i}\frac{4g\sqrt{J_1(J_1+1)} }{\omega_{A,J}^2 - \Delta m^2} \\
\times{\cal D}^{J_2M_2}_{J_1M_1, JM\rho}[\Phi_1^{J_1M_2},\bar\Phi_1^{J_2M_2}]\,.
\end{multline}

The dynamics of the theory close to the bound can now be derived by solving the constraints. 
The only surviving contribution to the kinetic term comes from $\Phi_1$, whose angular momenta are moreover constrained by the condition $\tilde{m} = -m = J$. 
The nonrelativistic nature of the resulting dynamics arises from Eq.~\eqref{eq:Phi_const}, which relates the canonical momentum to the complex conjugate field, just like in a nonrelativistic field theory. 
For convenience, we  introduce a new field variable
\begin{equation}
\label{eq:mom_field_rel}
\bar\Phi_s \equiv \sqrt{2(1+s)}\Phi_1^{\tfrac{s}{2},-\tfrac{s}{2},\tfrac{s}{2}}
\end{equation}
 where $s$ is now integer valued and not to be confused with the $\SU{4}$ index which no longer appears below. The choice of normalization guarantees a canonical Dirac bracket, whose nontrivial entry is derived from Eq.~\eqref{eq:Phi_const}, suppressing matrix indices, 
 \begin{equation}
 \label{eq:dir_brack}
 \{\Phi_s,\bar\Phi_{s'} \} = i\delta_{ss'}\,.
 \end{equation}
  A similar notation for the Clebsch--Gordan coefficients is adopted later. 
 
We obtain for the quadratic piece of the effective Hamiltonian
\begin{equation}
H_0 = \tr\sum_{s\geq 0}(s+1) |\Phi_s|^2\,.
\end{equation}
Since $H_0 - S_1 - Q_1 = 0$ by construction, we can now obtain the interacting Hamiltonian in the decoupling limit \eqref{eq:limit} as
\begin{equation}
H_\text{int} =  \lim_{g^2 N \to 0} \frac{H - S_1 - Q_1}{g^2 N}\,.
\label{eq:nr_lim}
\end{equation}
We insert all constraints and
make use of the symmetry $(s_1,s_2) \leftrightarrow (s_3,s_4)$, as well as of the fact that interactions are nontrivial only for $s_1+s_3 = s_2 + s_4$ due to angular momentum conservation,
to write the Hamiltonian as
\begin{equation}
H_\text{int} = \frac{1}{4N}\tr\sum_{s_1,s_2\geq 0}\sum_{l\geq 0} V^{s_1,s_2}_{l}[\bar\Phi_{s_1},\Phi_{s_1 + l}][\bar\Phi_{s_2+l},\Phi_{s_2}]\,,
\end{equation}
with
\begin{multline}
V^{s_1,s_2}_{l} \equiv \sum_{JM} \Bigg(\frac{(2+2s_1+l)(2+2s_2+l)}{8J(J+1)}  {\cal C}^{s_1+l}_{s_1,JM} {\cal C}^{s_2+l}_{s_2,JM} \\
- \sum_{\rho = \pm1}\frac{2\sqrt{s_1(s_1+2)}\sqrt{s_2(s_2+2)}}{\omega_{A,J}^2 - (m-\tilde{m})^2} {\cal D}^{s_1+l}_{s_1, JM\rho}{\cal \bar D}^{s_2+l}_{s_2, JM\rho}\\
+ \frac{1}{2}{\cal C}^{s_1+l}_{s_1,J M}{\cal C}^{s_2+l}_{s_2,J M} \Bigg)\,.
\end{multline}
Here, the short-hand notation for ${\cal C}$ and ${\cal D}$ is exactly as in Eq.~\eqref{eq:mom_field_rel}.
Using their explicit form, Eqs.~\eqref{eq:C_exp} and \eqref{eq:D_exp}, the summation over $J$ can be performed. 
The individual terms in this expression are complicated and quite nontrivial to evaluate. However, their combination reduces to the strikingly simple answer
\begin{equation}
V^{s_1,s_2}_{l > 0}= \frac{2}{l} \,,
\end{equation}
while the contributions for $l = 0$ are proportional to the $\SU{N}$ singlet constraint and hence vanish on all physical states and field configurations.

We introduce the coupling $g_0$ as the analogue of the 't Hooft coupling after the decoupling limit by defining the total Hamiltonian $H \equiv H_0 + g_0 H_\text{int}$. 
Using the $\SU{N}$ charge density in Fourier space,
\begin{equation}
q_s \equiv \sum_{n \geq 0} [\bar\Phi_n,\Phi_{n+s}]\,.
\label{eq:qun_fourier}
\end{equation}
we find the result
\begin{equation}
\label{eq:H_class_fin}
H = \tr\left(\sum_{s\geq 0}(s+1) |\Phi_s|^2 + \frac{g_0}{2N}\sum_{s > 0} \frac{1}{s} |q_s|^2\right)\,.
\end{equation}
This is a nonrelativistic field theory that describes the effective dynamics of ${\cal N} = 4$ SYM near the $\SU{1,1}$ BPS bound. The global $\U{1}$ symmetry is evident, since phase rotations of $\Phi$ leave the Hamiltonian invariant. This is a manifestation of the nonrelativistic nature of the theory. The invariance of the interaction under $\SU{1,1}$ transformations can be shown by considering the representation of the $\SU{1,1}$ generators on $\Phi_s$, 
\begin{equation}
\begin{array}{c}\displaystyle
L_0 = \tr\sum_{m\geq 0}\left(m+\tfrac{1}{2}\right)|\Phi_m|^2\,,\\[5mm]\displaystyle
L_+ = (L_-)^\ast = \tr\sum_{m\geq 0}\left(m+1\right) \Phi_{m+1}^\dagger \Phi_{m}\,,
\end{array}
\end{equation}
satisfying $\{L_0,L_\pm\} = \pm i L_\pm$ and $\{L_+,L_-\} = 2iL_0$. All generators commute with the interaction part of $H$ on the singlet constraint surface $q_0 = 0$.

In fact, the presence of the singlet constraint implies that the $\SU{N}$ symmetry of Eq.~\eqref{eq:H_class_fin} remains gauged. Indeed, it 
can be conveniently written by introducing an auxiliary field $\Psi_s$
as
\begin{multline}
\label{eq:H_local}
H = \tr\sum_{s \geq 0} \bigg((s + 1)\bar\Phi_s \Phi_s + s \bar\Psi_s \Psi_s \\
+ \sqrt{\frac{g_0}{2N}} (\Psi_{s} \bar{q}_s+\bar\Psi_s q_s) \bigg)\,,
\end{multline}
if supplemented by the constraint $\Pi_\Psi = 0$ and keeping in mind the form Eq.~\eqref{eq:dir_brack} for the bracket of $\Phi_s$. Remarkably, $\Psi_s$ plays here the role of a temporal gauge field that automatically enforces both the singlet constraint and gives rise to the interactions. The gauge redundancy becomes manifest when considering the Lagrangian. If it were not for the condition $s \geq 0$, we could directly obtain Eq.~\eqref{eq:H_local} from an action of a local $1+1$-dimensional gauge theory. Note that $s \geq 0$ can be viewed as a chirality condition if one identifies $s$ as a momentum along a circle. We will discuss this intriguing emergence of lower-dimensional locality in a forthcoming work \cite{BaigueraHarmarkWintergerst}.

\section*{Quantization and the one-loop dilatation operator}

We now proceed to quantize Eq.~\eqref{eq:H_local} in order to complete the diagram in Fig.~\ref{fig:commd}. To this end, we replace the Dirac bracket Eq.~\eqref{eq:dir_brack} by commutators, $\{\cdot,\cdot\} \to i[\cdot,\cdot]$, where we have put $\hbar \equiv 1$. We introduce ladder operators $a_s \equiv \Phi_s$, $a_s^\dagger = \bar\Phi_s$ that 
obey canonical commutation relations, i.e. $[(a_r)^i_j, (a_s^\dagger)^k_l] = \delta^i_l\delta^k_j\delta_{rs}$. 
We directly promote Eq.~\eqref{eq:H_class_fin} to the quantum Hamiltonian,
\begin{equation}
\label{eq:H_q}
H_q = \tr\left(\sum_{s\geq 0}(s+1) a_s^\dagger a_s + \frac{g_0}{2N}\sum_{s > 0} \frac{1}{s} q_s^\dagger q_s\right)\,.
\end{equation}
We justify this choice by showing that it leads to a normal ordered form that is fully equivalent to the one-loop dilatation operator as originally derived in \cite{Beisert2004,Beisert2004a}. In fact, this defines a procedure that allows us to straightforwardly read off the one-loop dilatation operator in a given subsector from the nonrelativistic Hamiltonian.

Normal ordering gives rise to self-energy corrections, concretely
\begin{multline}
\sum_{l>0}\frac{1}{l}\tr(q_l^\dagger q_l) = \sum_{l>0}\frac{1}{l}\tr(\normord{q_l^\dagger q_l})\\
+ 2 N \sum_{n = 0}^\infty h(n) \tr(a_{n}^\dagger a_{n}) - 2 \sum_{n = 0}^\infty h(n) \tr(a_{n}^\dagger) \tr(a_{n})\,,
\end{multline}
with the harmonic numbers $h(n) = \sum_{k=1}^n \tfrac{1}{k}$. 
The above corrections can be equivalently written in terms of a renormalized four-point interaction. Exploiting the $\SU{N}$ singlet condition, 
one can through simple manipulations of the sums derive the interaction Hamiltonian
\begin{multline}
H_\text{int} = \frac{1}{4N} \sum_{m = 0}^\infty \sum_{k,k'=0}^{m} \tr\left(\normord{\left[a_{k'}^\dagger,a_k\right]\left[a_{m-k'}^\dagger, a_{m-k}\right]}\right)\\
\times \left(\delta_{k=k'}(h(k) + h(m-k))- \delta_{k\neq k'}\frac{1}{|k-k'|}\right)\,,
\label{eq:hamdil}
\end{multline}
where here the square brackets denote matrix commutators. 
The second line of Eq.~\eqref{eq:hamdil} is precisely the one-loop dilatation operator in the bosonic $\SU{1,1}$-sector \cite{Beisert2004,Beisert2004a} and we have thus discovered a complementary way of calculating the one-loop dilatation operator in a given subsector without explicitly evaluating loop diagrams. This completes our derivation of the diagram in Fig.~\ref{fig:commd}.

We end by noting that the quantization prescription that was hereby forced upon us is yet another hint at the fundamental nature of the quasi-local theory defined by Eq.~\eqref{eq:H_local}. In a certain sense, it corresponds to treating both $\Phi_s$ and $\Psi_s$ as fundamental quantum degrees of freedom and imposing normal ordering on Eq.~\eqref{eq:H_local}.

\section*{Conclusions and outlook}
We have derived a novel interacting nonrelativistic field theory from a near-BPS limit of ${\cal N} = 4$ SYM. The resulting theory has a global $\U{1}$ symmetry as well as $\SU{1,1}$ invariant interactions and consists of a dynamical complex chiral scalar field interacting with a nondynamical gauge boson. 
We have focused on the near-BPS limit associated with the BPS bound \eqref{su11_bound} to provide a proof of concept. Our results apply to any other BPS bound of ${\cal N}=4$ SYM \cite{HarmarkKristjanssonOrselli2007}, with some small subtleties when including fermions, which we will address in \cite{HarmarkWintergerst}.

Due to its nonrelativistic nature, our novel field theory can be studied explicitly at any coupling, and as such should provide important insight into the workings of the AdS/CFT correspondence. 
The limit that we propose can be taken directly in string theory, giving rise to nonrelativistic string theories on  $\U{1}$-Galilean  target space \cite{HarmarkKristjanssonOrselli2009,HarmarkHartongObers2017,HarmarkHartongMenculiniObersYan2018,HarmarkHartongMenculiniObersOling2019}\footnote{See \cite{Kluson2018,BergshoeffGomisYan2018} for related work.} and D-branes \cite{Harmark2016}.
Indeed, our work immediately opens up the route to reexamine many renowned features of holography. This begins with studying details of the emergence of bulk geometry, for example from the entanglement structure \cite{VanRaamsdonk2010}. Considering other BPS bounds can elucidate the question of how and in particular how many \cite{AldayPerlmutter2019} additional dimensions are encoded in the theory, since they are expected to be dual to bulk configurations of different dimension \cite{HarmarkHartongObers2017,HarmarkHartongMenculiniObersYan2018}. A thermal analysis can shed light on the precise details of the confinement/deconfinement transition and allows for a quantitative study of the recently suggested mechanism of partial deconfinement \cite{HanadaIshikiWatanabe2019,HanadaJevickiPengWintergerst2019}. Studying temperatures above the Hawking--Page transition is particularly interesting for the near-BPS limit with $SU(1,2|3)$ symmetry, which is expected to contain black holes \cite{GutowskiReall2004}. Consequently, it should for example exhibit maximal chaos \cite{MaldacenaShenkerStanford2016}. 
 The explicit construction presented in this note will allow one to explore
the nonrelativistic corners
  of holography in quantitative detail. 

Aside from the possible application for holography, finding new nonrelativistic field theories from near-BPS of ${\cal N}=4$ SYM is interesting in its own right. This points to a  family of novel nonrelativistic quantum field theories, some with supersymmetry, whose properties have yet to be explored.

\begin{acknowledgments}
We are grateful to Jelle Hartong and Marta Orselli for discussions and Gerben Oling for discussions and detailed comments on the manuscript. This work was supported by FNU grant number DFF-6108-00340.
\end{acknowledgments}

\bibliography{bibliography}

\end{document}